\documentclass{optica-article}

\journal{opticajournal} 

\articletype{Research Article}

\usepackage{hyperref}
\usepackage{url}

\begin{document}

\title{Optical design for implementing non-collinear circularly polarized high harmonic generation in an enhancement cavity}

\author{P.\,DAVE, T.\,AUGUSTE, O.\,GIRARD, T.\,RUCHON, and D.\,BRESTEAU\authormark{*}}

\address{Université Paris-Saclay, CEA, LIDYL, 91191 Gif-sur-Yvette, France}

\email{\authormark{*}david.bresteau@cea.fr} 


\begin{abstract*} 
We introduce an original optical cavity design intended to efficiently output-couple extreme ultraviolet (XUV) light produced via cavity-enhanced high harmonic generation (CE-HHG). It supports the amplification of oppositely circularly polarized modes, crossing with a small angle in the high finesse cavity, where circularly polarized HHG scheme (NCP-HHG) \cite{Hickstein_Dollar_Grychtol_Ellis_Knut_Hernández-García_Zusin_Gentry_Shaw_Fan_et_al._2015} will be implemented. We present an analytical model, numerical simulations, and experimental results obtained with a low-power continuous (CW) laser, demonstrating the strong potential of the cavity as an efficient XUV output coupling method. In particular, it is well suited for producing the 7th harmonic (H7) from an Ytterbium frequency comb at 8.4\,eV, which is of particular interest for precision nuclear spectroscopy of Thorium.
\end{abstract*}


\section{Introduction}
High harmonic generation (HHG) has emerged as an essential tool for attosecond science, producing coherent extreme ultraviolet (XUV) radiation from tabletop sources. Since its discovery \cite{Ferray_L’Huillier_Li_Lompre_Mainfray_Manus_1988}, HHG has enabled advances in probing electronic dynamics in multiple phases of matter. The most standard and simple way to design a gas-phase HHG source is called single-pass design. A femtosecond mode-locked oscillator is amplified through a regenerative amplifier to reach a pulse energy on the order of 1\,mJ. The pulse is then strongly focused into a gas target under vacuum where the peak intensity reaches $10^{14}$ W/cm². The HHG process produces XUV light with an energy conversion efficiency of the order of $10^{-6}$. The driving pulse is then mostly unaffected by the process, but is dumped after this single interaction. Single-pass HHG sources are then limited by the decrease in repetition rate in the regenerative amplifier to roughly <1\,MHz.

The idea of the CE-HHG technique \cite{Gohle_Udem_Herrmann_Rauschenberger_Holzwarth_Schuessler_Krausz_Hänsch_2005,Jones_Moll_Thorpe_Ye_2005} is to put the generation medium inside an optical cavity that recycles the pulses. By locking the cavity of the oscillator with this external cavity, the injected pulses can be coherently superimposed, which lead to a passive power enhancement. This gain in efficiency allows to reach much higher repetition rates, since all the pulses from the mode-locked oscillator are amplified, which dictates a repetition rate of the order of 100\,MHz. The drawback is that the demand on the phase locking of the laser is much higher. CE-HHG founds its most successful applications with time-resolved photo-emission spectroscopy \cite{Allison2025}, and recently opened the field of precise spectroscopy of nuclear transitions \cite{Zhang2024}. Indeed CE-HHG is currently the only demonstrated technique to allow the up-conversion of an infrared frequency comb into the XUV domain. However, efficient XUV output coupling from the enhancement cavity remains a challenge, and existing schemes such as holed mirrors, Brewster plates, or gratings reduce the finesse and limit the output coupling efficiency to tens of percent \cite{Pupeza_Zhang_Högner_Ye_2021}.

\begin{figure}[!htb]
    \centering
    \includegraphics[width=1.0\linewidth]{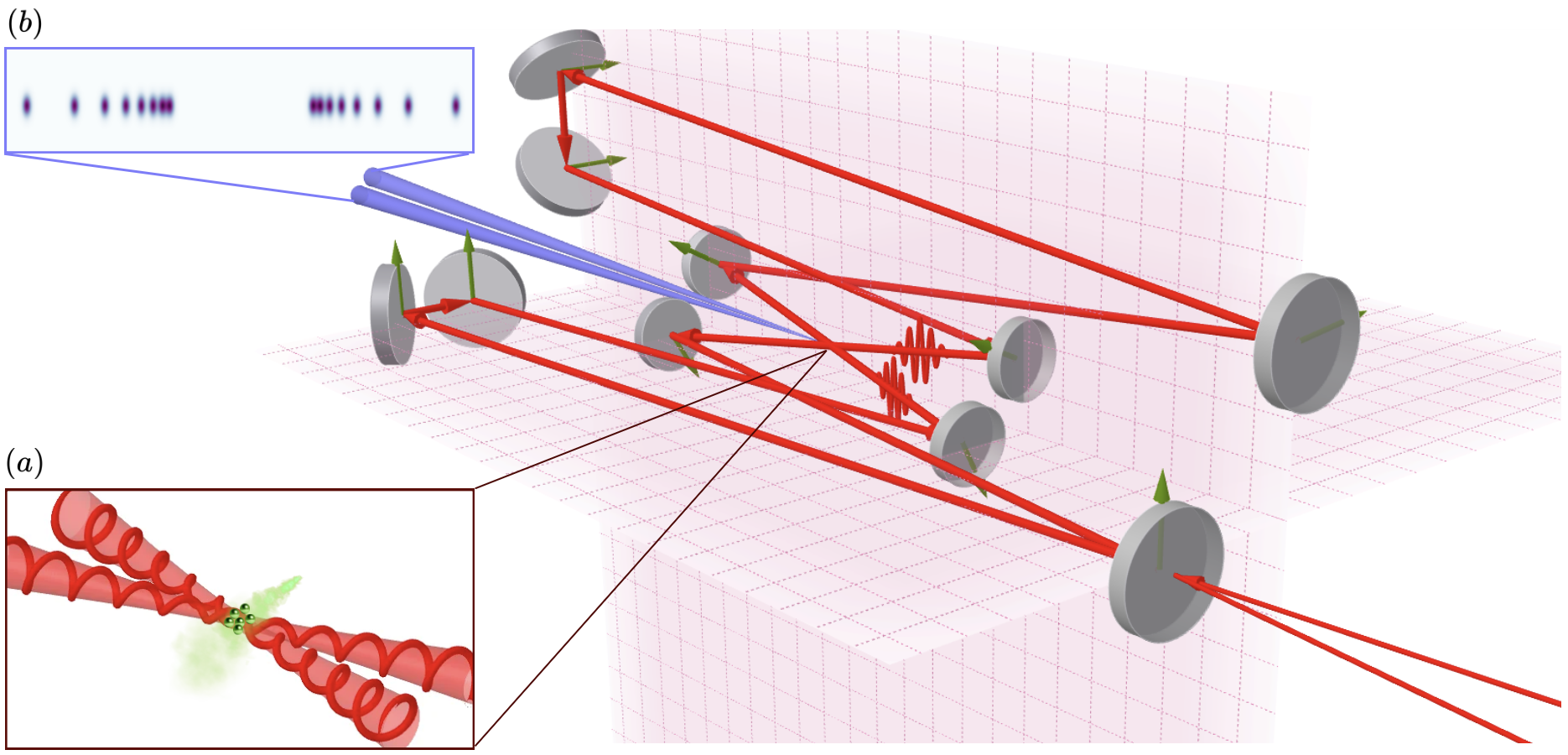}
    \caption{\textbf{Implementing NCP-HHG inside a cavity:} Overview of the all-reflective cavity design satisfying the constraints outlined in the text. (a) Close-up of the interaction region showing the two counter-propagating pulses of opposite handedness ($\sigma^+$, $\sigma^-$) crossing at the focus and (b) the resulting XUV beamlets for all harmonic orders q emitted at $\pm \Theta/q$ from the bisector.}
    \label{fig:overall}
\end{figure}

\begin{figure}[!htb]
    \centering
    \includegraphics[width=1.0\linewidth]{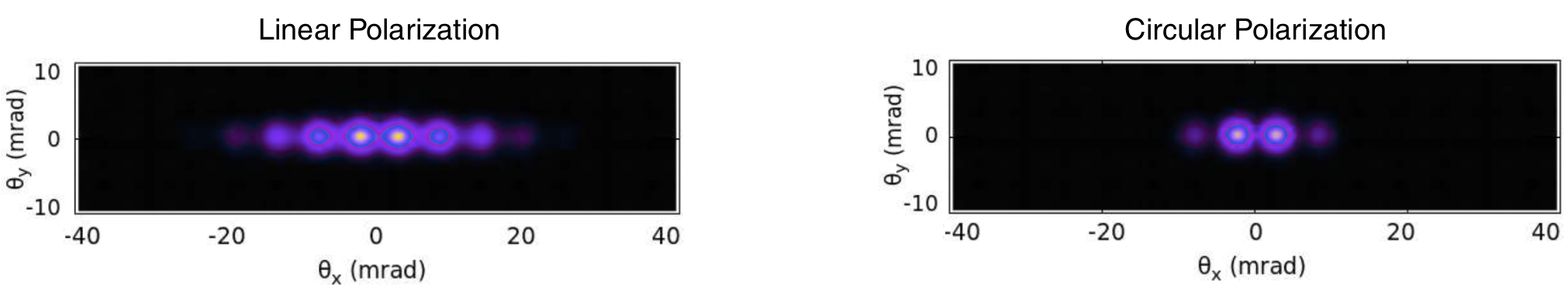}
    \caption{\textbf{Numerical simulations of the HHG generation and XUV light propagation} to compare the NCP-HHG scheme (right) and the generation with linearly polarized light in the same conditions (left). Both signals are normalized to their maximum. The 15th harmonic generated with the HHG process is simulated using the non-adiabatic paraxial wave equation with a dipole source term calculated in the strong field approximation (SFA) in argon with a driver wavelength of 1030\,nm, a focal length of 100\,mm, a pulse energy of 10\,$\mu$J, a pulse duration of 100\,fs, a beam waist radius at focus of 30\,$\mu$m, and a target length of 100\,$\mu$m. The horizontal axis corresponds to the angle of emission in the plane formed by the two IR driving beams. The full angle between the driving beams is 80\,mrad.}
    \label{fig:sim_sfa}
\end{figure}

We propose a new output coupling method based on a non-collinear HHG scheme, in which two driving beams propagating at a small angle intersect at their focal point. The basic idea being that, thanks to the conservation of momentum, the XUV will be mainly emitted close to the bisector of the two IR beams, in a different direction to the driving beams, and can be efficiently output coupled into the gap between the two subsequent mirrors, as illustrated in Fig.~\ref{fig:overall}. More precisely, denoting \(\mathbf{k}_1, \mathbf{k}_2\) the wave-vectors of the drivers, emission occurs along all directions \(\Theta\) of harmonic \(q\) such that \(\mathbf{k}_q=n_1\mathbf{k}_1+n_2\mathbf{k}_2\), where \(n_1,n_2 \in \mathbb{N}\) and \(n_1+n_2=q\). This scheme has already been successfully implemented with linearly polarized light \cite{Zhang_Schoun_Heyl_Porat_Gaarde_Ye_2020}, and still inspire some further work \cite{PhysRevResearch.7.023071}. Unfortunately, with linear polarization, a standing-wave intensity grating forms at focus, generating multiple diffraction orders \cite{Chappuis2019} that are detrimental to the output coupling efficiency. In order to overcome this difficulty, we propose the use of two counter-rotating circularly polarized pulses to drive HHG, a scheme known as NCP-HHG \cite{Hickstein_Dollar_Grychtol_Ellis_Knut_Hernández-García_Zusin_Gentry_Shaw_Fan_et_al._2015}, to suppress these high diffraction orders by virtue of the conservation of the photon spin. This should concentrate even further the XUV emission close to the bisector and allow the use of larger angles at crossing. The comparison of both situations is illustrated by numerical simulations in Fig.~\ref{fig:sim_sfa}. Although demonstrated in a single-pass geometry, this scheme has not yet been realized within an enhancement cavity. Implementing this configuration imposes stringent design constraints reviewed in Sec.~\ref{sec:design_constraints}, supported by numerical simulations of the HHG emission and propagation. In Sec.~\ref{sec:jones_3d}, we present a theoretical formalism to predict the polarization states stored in a 3D optical cavity. We show in particular that non-planar cavity geometries are required to sustain circularly polarized eigenmodes. This formalism is applied to the proposed design in Sec.~\ref{sec:spacecraft_theory}, demonstrating that it satisfies all requirements for NCP-HHG in cavity. In Sec.~\ref{sec:spacecraft_experiment} we present an experimental implementation using a low-power CW laser and compare the results with theoretical predictions. Finally, Sec.\ref{sec:femtosecond_enhancement} addresses the extension to femtosecond enhancement and CE-HHG.

\section{Design Constraints for the implementation of NCP-HHG in an Enhancement Cavity} \label{sec:design_constraints}

Inside a femtosecond enhancement cavity at resonance, and with a perfect coupling, the average optical power $P_\mathrm{cav}$ is given by
\begin{equation}
P_\mathrm{cav} = G P_\mathrm{in},
\end{equation}
where $G$ is the power enhancement factor and $P_\mathrm{in}$ the input mean power. In the NCP configuration, two identical pulses are circulating simultaneously and share the intracavity power, giving a pulse energy of
\begin{equation}
\mathcal{E}_\mathrm{pulse} = \frac{G P_\mathrm{in}}{2 f_\mathrm{rep}},
\end{equation}
with $f_\mathrm{rep}$ the driving laser's repetition rate.

For a pulse with gaussian spatial and temporal profiles focusing with radius waist $w_0$ and pulse duration $\tau$, the peak intensity is
\begin{equation}
I_\mathrm{peak} = \frac{2\mathcal{E}_\mathrm{pulse}}{\pi w_0^2 \tau}
= \frac{GP_\mathrm{in}}{\pi w_0^2 \tau f_\mathrm{rep}}.
\end{equation}

Reasonable parameters $P_\mathrm{in}=10$\,W, $G=1000$, $\tau=100$\,fs, $f_\mathrm{rep}=100$\,MHz, and $w_0\simeq30\,\mu$m, yield $P_\mathrm{cav}\approx10$\,kW, $\mathcal{E}_\mathrm{pulse}\approx50\,\mu$J, and $I_\mathrm{peak}\approx4\times10^{13}$\,W/cm$^2$, which is in the range of peak intensities required to trigger the HHG process. They will be used through out unless otherwise stated. To successfully implement the NCP-HHG scheme, the cavity geometry must satisfy several stringent constraints:

\textbf{(i) All-reflective design.} A major advantage of the non-collinear generation scheme is that it proposes an XUV extraction strategy without deterioration of the finesse of the cavity, which should allow a better power enhancement. The use of intracavity transmissive elements is hereby avoided.

\textbf{(ii) Synchronization constraints.} Coherent superposition requires that the cavity round-trip time matches a multiple of the time separation between two pulses. In the NCP case, two pulses should be stored simultaneously, which leads to
\begin{equation}
L_\mathrm{rt} = \frac{2c}{f_\mathrm{rep}},
\end{equation} which corresponds to a 6m round-trip cavity length. For successive pulses to overlap temporally at the spatial crossing of the two beams, the half round-trip should be precisely 3\,m.

\textbf{(iii) Tight focusing, cavity length and damage threshold.} As derived above, achieving the peak intensity necessary to trigger the HHG process requires relatively tight focusing. The condition (ii) can only be met if the distance between the focusing mirrors is independent of the total cavity length, which rejects the possibility of a two-mirror cavity.

Another aspect to take into account is the optical fluence on the cavity mirrors, given by
\begin{equation}
F_m = \frac{\mathcal{E}_\mathrm{pulse}}{\pi w_m^2},
\end{equation}
which must remain below the coating damage threshold, requiring large enough beam radii $w_m$ on the mirrors. We implement this with spherical mirrors of focal length $f=+100$\,mm, producing a waist radius $w_0 \approx 30\,\mu$m at two spatially overlapped foci in the cavity center. In those conditions the waist radius at the focusing mirror surfaces is about $w_m = 1$\,mm, so the fluence is about $F_m \approx 1\,\mathrm{mJ/cm^2}$, which is far below the best damage thresholds specified by the mirror suppliers (\(\gtrsim 500 \,\mathrm{mJ/cm^2}\)).

\textbf{(iv) Spatial overlap of the two foci and XUV extraction.} To achieve the non-collinear generation, the two foci should cross with a half-angle $\Theta$, with a typical value of 40\,mrad. In this geometry, harmonic radiation is emitted predominantly along discrete directions close to the bisector, with the $q$-th harmonic emitted at angle $\Theta_q \approx \Theta/q$ \cite{Hickstein_Dollar_Grychtol_Ellis_Knut_Hernández-García_Zusin_Gentry_Shaw_Fan_et_al._2015}. This formula and the numerical simulations (Fig.~\ref{fig:sim_sfa}) predict an half-angle of emission of about 13.3\,mrad for H3. Therefore, a separation of 3mm of the inner edges of the subsequent mirrors (about 0.1m from the source point) should be reasonable to efficiently output couple H3 - and all the higher harmonic orders - while ensuring a proper reflection of the two driving beams.

\textbf{(v) Polarization eigenmodes.} The cavity must support counter-rotating circular eigenmodes ($\sigma^+$ and $\sigma^-$) at the two foci. This is probably the toughest constraint when associated to constraint (i), since it precludes the use of waveplates. First, it is needed that the cavity stores circularly polarized states, which implies to consider non-planar geometries, as will be explained in the following section. Secondly, a strategy should be employed to switch the helicity of the pulses for every half round-trip. To ensure this feature, an odd number of reflections per half round-trip enforces opposite handedness at the interaction region, which constrains the total mirror count to $n = 4k+2$, with $k \in \mathbb{N}$. For practical reasons, we used $k=2$ and a total of 10 mirrors.

\section{Jones Matrix Formalism for 3D Cavities} \label{sec:jones_3d}

In order to calculate the polarization of the resonant states of light in each segment of a 3D cavity as a function of its geometry, we use an extension of the Jones formalism \cite{Jones_1941}. It consists in inserting a rotation matrix between all pairs of successive mirrors matrices to perform a change of basis between the output of the first mirror and the input of the second one \cite{Chow_Gea-Banacloche_Pedrotti_Sanders_Schleich_Scully_1985, Nilsson1989, Zomer_Fedala_Pavloff_Soskov_Variola_2009}. In this section we introduce the general formalism and our notations. It will be useful to present the \textit{Spacecraft} design in the next section. The \textit{GeoGebra} free software \cite{geogebraGeoGebraWorlds} is a precious tool to ease the representation of the 3D geometry.

\begin{figure}[!htb]
    \centering
    \includegraphics[width=1.0\linewidth]{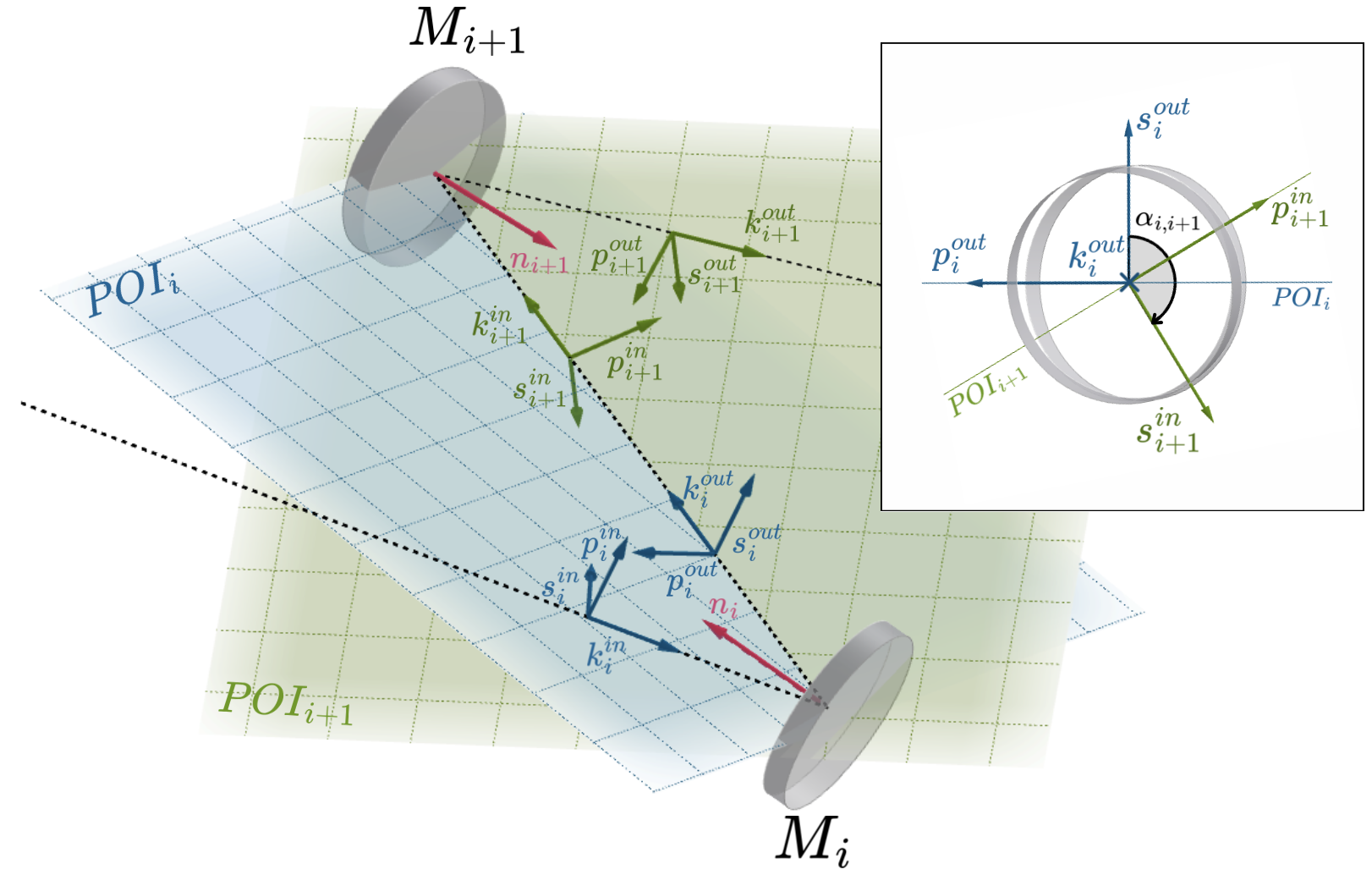}
    \caption{\textbf{3D Mirror Basis:} Illustration of the incoming and outgoing basis vectors for two successive mirrors $M_i$ and $M_{i+1}$ oriented in arbitrary directions. The intersection of the planes of incidence $\text{POI}_i$ and $\text{POI}_{i+1}$, associated with each reflection, correspond to the beam propagation from $M_i$ to $M_{i+1}$. \textit{Inset:} Illustration of the same bases seen from behind mirror $M_{i}$, looking in the direction where the laser is travelling. From this perspective, $\alpha_i$ is defined as the angle from $\mathbf{s}_i^{\text{out}}$ to $\mathbf{s}_{i+1}^{\rm in}$ with absolute value inferior to $\pi$. It is defined positive if it is counterclockwise (therefore negative in this inset). Here is the associated \href{https://www.geogebra.org/m/w5w6df5s}{GeoGebra file}.}
    \label{fig:basis}
\end{figure}

We consider an $n$-mirror cavity with the active surfaces of mirrors intercepting the beam at positions $\mathbf{m}_i = (x_i, y_i, z_i)$, with indices taken cyclically. At each mirror $i$, we introduce a local right-handed polarization basis for the incoming $B_i^{\rm in}= \{ \mathbf{k}_i^{\rm in}, \mathbf{p}_i^{\rm in}, \mathbf{s}_i^{\rm in}\}$ and outgoing $B_i^{\text{out}}= \{ \mathbf{k}_i^{\text{out}}, \mathbf{p}_i^{\text{out}}, \mathbf{s}_i^{\text{out}}\}$ beam, as shown in Fig.~\ref{fig:basis}:
\newline

\begin{minipage}{0.48\textwidth}
\begin{equation}
 B_i^{\rm in} =  \left\{
\begin{aligned}
\mathbf{k}_i^{\rm in} &=\frac{\mathbf{m}_i - \mathbf{m}_{i-1}} {\left\lVert \mathbf{m}_i - \mathbf{m}_{i-1} \right\rVert} \\[6pt] \mathbf{p}_i^{\rm in} &=  \mathbf{s}_i^{\rm in} \times \mathbf{k}_i^{\rm in} \\[6pt]
\mathbf{s}_i^{\rm in} &=
\frac{\mathbf{k}_i^{\rm in} \times \mathbf{n}_i}
{\left\lVert \mathbf{k}_i^{\rm in} \times \mathbf{n}_i \right\rVert}
\end{aligned}
\right.
\end{equation}
\end{minipage}
\hspace{0.5cm}
\begin{minipage}{0.48\textwidth}
\begin{equation}
B_i^{\text{out}} = \left\{
\begin{aligned}
\mathbf{k}_i^{\text{out}} &=\frac{\mathbf{m}_{i+1} - \mathbf{m}_{i}}{\left\lVert \mathbf{m}_{i+1} - \mathbf{m}_{i} \right\rVert} \\[6pt]\mathbf{p}_i^{\text{out}} &=  \mathbf{s}_i^{\text{out}} \times \mathbf{k}_{i}^{\text{out}} \\[6pt]\mathbf{s}_i^{\text{out}} &= \mathbf{s}_i^{\rm in}
\end{aligned}
\right.
\end{equation}
\end{minipage} where the surface normal is taken as the normalized bisector of the incoming and
outgoing propagation directions,
\(
\mathbf{n}_i =
(-\mathbf{k}_i^{\rm in} + \mathbf{k}_i^{\text{out}}) /
\lVert -\mathbf{k}_i^{\rm in} + \mathbf{k}_i^{\text{out}} \rVert
\).

In order to switch from the expression of the components of the field in the local basis of mirror $i$ to the local basis of mirror $i+1$ we use the rotation matrix $\mathbf{R}_{i}$, defined as follow:

\begin{equation}
\begin{pmatrix}
E_{i+1}^{\rm in, p} \\
E_{i+1}^{\rm in,s}
\end{pmatrix}
= \mathbf{R_{i}}
\begin{pmatrix}
E_{i}^{\text{out},p} \\
E_{i}^{\text{out},s}
\end{pmatrix}
= 
\begin{pmatrix}
\mathbf{p}_i^{\text{out}}.\mathbf{p}_{i+1}^{\rm in} & \mathbf{s}_i^{\text{out}}.\mathbf{p}_{i+1}^{\rm in} \\
\mathbf{p}_i^{\text{out}}.\mathbf{s}_{i+1}^{\rm in} & \mathbf{s}_i^{\text{out}}.\mathbf{s}_{i+1}^{\rm in} \\
\end{pmatrix}
\begin{pmatrix}
E_{i}^{\text{out},p} \\
E_{i}^{\text{out},s}
\end{pmatrix}
\label{eq:rotproj}
\end{equation}

At this stage, the introduction of an angle is not strictly necessary. In particular, our numerical simulations do not rely on it, and only need the projection relation given by Eq.\,(\ref{eq:rotproj}). However, the introduction of an angle provides a better intuition with rotation transformations. The sign convention for these angles requires careful definition, as successive rotations will be considered in the following section. A practical definition of the angle is illustrated in the inset of Fig.\,\ref{fig:basis} given as follow: looking from behind mirror $M_i$ in the direction $\mathbf{k}_i^{\text{out}}=\mathbf{k}_{i+1}^{\rm in}$ where the beam is travelling, $\alpha_i$ is defined as the angle from $\mathbf{s}_i^{\text{out}}$ to $\mathbf{s}_{i+1}^{\rm in}$ with absolute value inferior to $\pi$. It is defined positive if it is counterclockwise.

With this definition, we have \cite{Nilsson1989}

\begin{equation}
\mathbf{R}_i = \mathbf{R}(\alpha_i) =
\begin{pmatrix}
\cos \alpha_i & -\sin \alpha_i \\
\sin \alpha_i & \cos \alpha_i
\end{pmatrix}
\label{eq:alphadefinition}
\end{equation}

Reflection at mirror $i$ is described by the Jones matrix
\begin{equation}
\mathbf{M}_i =
\begin{pmatrix}
\rho_i^p e^{i\phi_i} & 0 \\
0 & \rho_i^s e^{-i\phi_i}
\end{pmatrix}
\label{eq:mirrorrefl}
\end{equation}
with $2\phi_i = \phi_i^p - \phi_i^s = \Delta \phi$ where $\rho_{i}^{p,s}$ and $\phi_{i}^{p,s}$ denote respectively the reflectivities and phase shifts for the p- and s-polarization. Note that the input vector is expressed in the $B_i^{\rm in}$ basis, while the output one is expressed in the $B_i^{\text{out}}$ one. The $\pi$ phase shift acquired at each reflection is not included in the definition of the phases $\phi_i^p$ and $\phi_i^s$, we rather introduce it explicitly with the introduction of the reflection operator \cite{Nilsson1989} at each reflection

\begin{equation}
\mathbf{T} =
\begin{pmatrix}
-1 & 0 \\
0 & 1
\end{pmatrix}.
\end{equation}

As shown in the Appendix \ref{sec:theory_fabry_perot}, calculating the polarization state of the resonant field in the $\Xi_i$ segment of the cavity from mirror $M_i$ to $M_{i+1}$ reduces to the mathematical problem of diagonalizing the round-trip matrix

\begin{equation}
\mathbf{J}_i =\mathbf{M}_i\mathbf{T}\mathbf{R}(\alpha_{i-1})\mathbf{M}_{i-1}\cdots\mathbf{M}_0\mathbf{T}\mathbf{R}(\alpha_{n-1})\mathbf{M}_{n-1}\cdots\mathbf{M}_{i+1}\mathbf{T}\mathbf{R}(\alpha_{i}),
\label{eq:roundtrip_matrix}
\end{equation} expressed in the $B_i^{\text{out}}$ basis. Note that by convention we start the indexing of the mirrors at $i=0$ with $M_0$ being the input coupler (IC).

The matrix $\mathbf{J}_i$ describes the polarization transformation accumulated over one cavity round trip, starting and ending at the mirror $i$. We denote its complex
eigenvalues $\mu^{\pm}$ and eigenvectors
$\lvert \mathbf{u}_{i}^{\pm} \rangle$ satisfy
\begin{equation}
\mathbf{J}_i \lvert \mathbf{u}_{i}^{\pm} \rangle
=
\mu^{\pm} \lvert \mathbf{u}_{i}^{\pm} \rangle.
\end{equation}

The eigenvectors $\lvert \mathbf{u}_{i}^{\pm} \rangle$ define the polarization
eigenmodes of the cavity in section $i$ and are the only polarization states that
resonate in that section. Expressed in the $B_i^{\text{out}}$ basis, they take the form
\begin{equation}
\lvert \mathbf{u}_i \rangle =
\begin{pmatrix}
u_i^p \\ u_i^s
\end{pmatrix},
\end{equation} where $u_i^p$ and $u_i^s$ are the complex field amplitudes along the local $p$ and $s$
axes, respectively.

The eigenvalues encode both round-trip loss $|\mu^{\pm}|$ and accumulated phase $\arg(\mu^{\pm})$, with their phase difference resulting in polarization-dependent resonance splitting. Notice that they do not depend on the segment considered (see Eq.~(\ref{eq:transformeigvecs})), which makes sense since they are related to global properties of the cavity.

To quantify the polarization state of an eigenmode, we compute the Stokes
parameters from the Jones vector components $(u_i^p,u_i^s)$. The normalized degree of circular polarization is given by:
\begin{equation}
\left|s_3\right|=\left|\frac{S_3}{S_0}\right|=\frac{2\,\left|\mathrm{Im}\!\left(u_i^{p*} u_i^s\right)\right|}
{|u_i^p|^2 + |u_i^s|^2},
\label{eq:s3_expression}
\end{equation}
which ranges from $0$ for linear polarization to $1$ for purely circular
polarization, with intermediate values corresponding to elliptical states. This metric provides a convenient scalar measure of circularity independent of
handedness, allowing us to directly evaluate and optimize 3D cavity geometries for circular polarization eigenmodes at selected sections inside the cavity.

\subsection{Planar Cavities and Linear Eigenmodes}

In a planar n-mirror cavity, each mirror contributes as a reflection matrix in the form of Eq.(\ref{eq:mirrorrefl}). Since all mirrors share a common plane of incidence, the rotation matrices reduce to the identity, and the round-trip Jones matrix is simply the ordered product of mirror matrices:
\begin{equation}
    \mathbf{J}=\prod_{i=0}^{n-1} \mathbf{M}_i =  \begin{pmatrix}
    \prod_{i=0}^{n-1} \rho_i^p e^{i\phi_i} & 0 \\ 0 & \prod_{i=0}^{n-1} \rho_i^s e^{-i\phi_i}
    \end{pmatrix}.
\end{equation}

Because $\mathbf{J}$ is diagonal for any choice of mirror parameters, its eigenvectors are fixed by the matrix structure alone, independently of the values of \(\rho_i^{p,s} \) and \(\phi_i\):

\begin{equation}
    \lvert \mathbf{u}^{+} \rangle = \begin{pmatrix}
    1 \\ 0
    \end{pmatrix}, \quad \lvert \mathbf{u}^{-} \rangle = \begin{pmatrix}
    0 \\ 1
    \end{pmatrix},
\end{equation}
corresponding to pure p- and s-polarised eigenmodes, respectively. The sole exception arises when the two eigenvalues become degenerate, \( \prod_{i=1}^n \rho_i^p e^{i\phi_i} = \prod_{i=1}^n \rho_i^s e^{-i\phi_i}\). Under these conditions, the Jones matrix is proportional to the identity, so that every polarisation state is a degenerate eigenmode. However, this mathematical singularity is very difficult to achieve in experiments where any deviation from either condition (owing to mirror or angle imperfections) immediately lifts the degeneracy and restores the linear eigenmodes. A possible exception may be the 2-mirror case with a moderate finesse.

Thus, in planar optical cavities operated at non-normal incidence, no matter the size or configuration, the geometry fixes a global $s$ and $p$ polarization axes along the optical path, so that only linearly polarized eigenmodes are supported. However, circular polarization eigenmodes can be realized in 3D cavity geometries, where successive coordinate rotations between mirrors eliminate any globally privileged plane.

\section{\textit{Spacecraft} cavity design} \label{sec:spacecraft_theory}

\begin{figure}[!htb]
    \centering
    \includegraphics[width=1.0\linewidth]{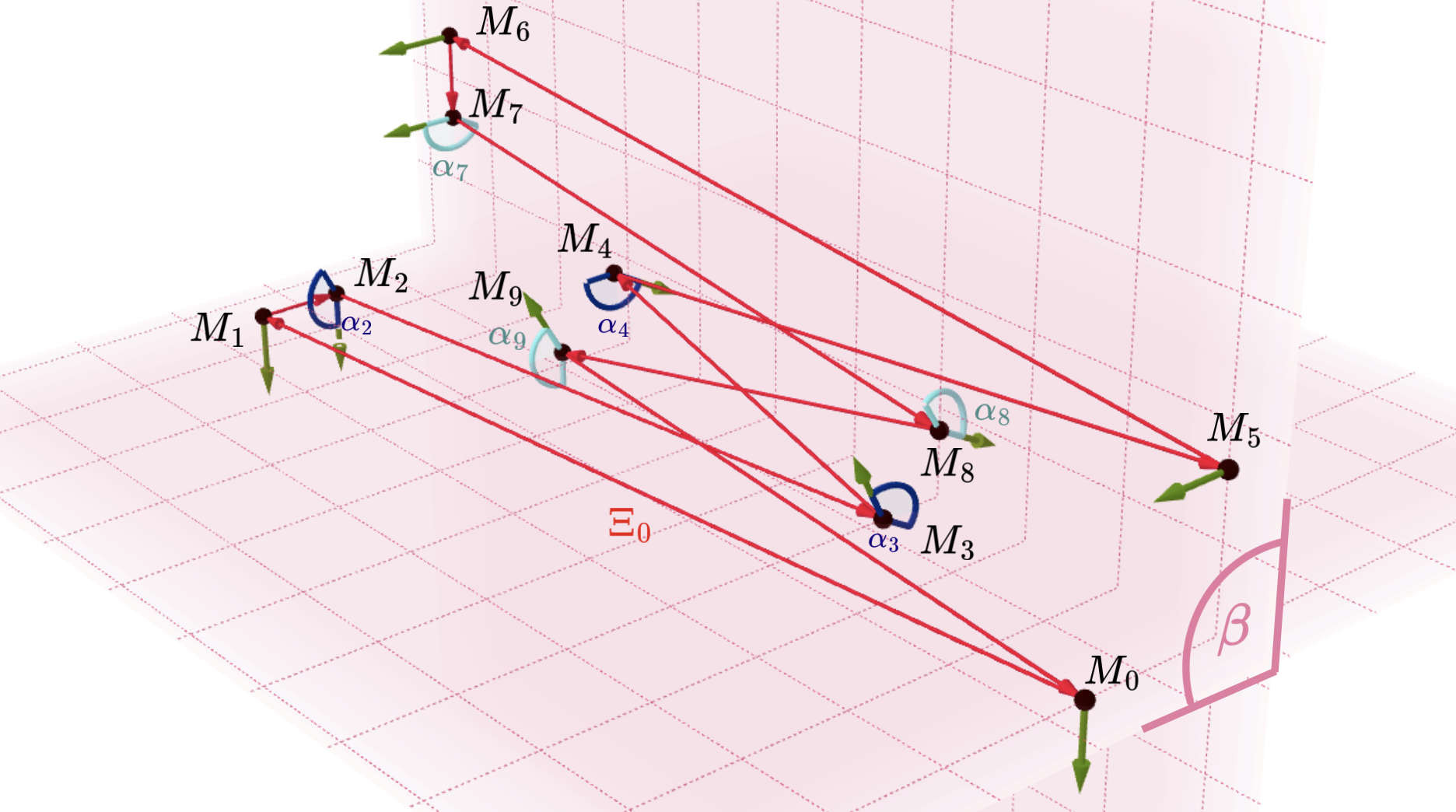}
    \caption{\textbf{Spacecraft cavity geometry:} The numbering of the mirrors from $M_0$ to $M_9$ is ordered according to the propagation of the beam, with $M_0$ being the IC. The $\alpha_i$ angles are defined by Eq.~(\ref{eq:alphadefinition}). The angles in dark blue and light blue correspond to series of rotations that complete the transformation from one plane to another, with their sum equal to the angle between the planes (\(\beta\)). The green arrows represents the $\mathbf{s}_i$ vectors, orthogonal to the POI at each mirror. Here is the associated \href{https://www.geogebra.org/3d/kx9vmwyp}{Geogebra file}.}
    \label{fig:points}
\end{figure}

In this section we present a particular cavity geometry that respects all the criteria exposed in Sec. \ref{sec:design_constraints}. We call it the \textit{Spacecraft} design. It is represented in Fig.~\ref{fig:points}. We then use the formalism exposed in Sec.~\ref{sec:jones_3d} to demonstrate that its eigenmodes are circularly polarised.

The qualitative reasoning that led to this geometry is the following. The constraint (iv) necessarily imposes that 4 focusing mirrors be placed in a common plane to produce the spatial intersection of the foci, which constitutes the central part of the cavity around the generation point. By performing the full cavity's stability analysis with the \textit{reZonator} software \cite{reZonator}, we find that the beam radius at focusing mirrors is approximately 1\,mm, which for the 40\,mrad half-angle configuration results in a 6\,mm maximum separation of the inner edges of the mirrors, i.e. twice the required spacing to output-couple H3. After passing through one of the focal point, the beam should undergo an odd number of reflections before reaching the second focus, to implement the helicity switching imposed by the constraint (v). Since transferring the beam from "one side of the cavity to the other" with an odd number of reflections is not possible using only low-AOI mirrors, two mirrors at $45^\circ$ AOI are introduced in each half of the cavity. The resulting geometry therefore comprises four high-AOI mirrors and six low-AOI mirrors. Finally, since the cavity must store circularly polarized light, it seems necessary to nullify any phase difference that is imposed between the local $s$ and $p$ components of the transverse field at each reflection. For this reason the cavity is split into two symmetric \textit{wings} - each one constituted of 5 mirrors - placed in orthogonal planes. The wings are symmetric in the sense that they are composed of the same sequence of mirrors reflecting at the same AOI. The idea behind this is as follows. If one consider for example the $p$ component of the field propagating in the vertical wing (composed of mirrors $M_4$, $M_5$, $M_6$, $M_7$, $M_8$), it undergoes a dephasing corresponding to a $p$ reflection at each mirror. Then in the central part of the cavity this component is rotated by 90° thanks to the geometric arrangement of the mirrors, just as a well oriented half-wave plate would do. In the horizontal wing (composed of mirrors $M_9$, $M_0$, $M_1$, $M_2$, $M_3$), the component undergoes a dephasing corresponding to an $s$ reflection on each mirror, which exactly compensates the accumulated dephasing in the horizontal wing. Therefore in a full round-trip, the dephasing is null. This qualitative reasoning is demonstrated more rigorously in the following. We will be particularly interested in finding the conditions for having $\sigma^+$ and $\sigma^-$ simultaneously in the two focusing segments of the cavity ($M_3M_4$ and $M_8M_9$). 

\subsection{Simplified Analytical Model}
\label{sec:analytical}

The round-trip matrix for this 10-mirror cavity in segment $\Xi_0$ is given by:
\begin{equation}
    \mathbf{J}_0 = \mathbf{M}_0\mathbf{T}\mathbf{R}(\alpha_9)\mathbf{M}_9\mathbf{T}\mathbf{R}(\alpha_8)\mathbf{M}_8 \cdots \mathbf{M}_2\mathbf{T}\mathbf{R}(\alpha_1)\mathbf{M}_1\mathbf{T}\mathbf{R}(\alpha_0)
\end{equation}
To pursue our analytical analysis, we introduce two simplifying 
assumptions. 

\begin{enumerate}
    \item The two high-AOI reflections at mirror pairs 1/2 and 6/7 occur in the same plane of incidence and therefore involve no rotation, and can be approximated with a single mirror matrix:
\begin{equation}
\mathbf{M} \sim \mathbf{M}_2\mathbf{T}\mathbf{R}(\alpha_1)\mathbf{M}_1\mathbf{T}\mathbf{R}(\alpha_0) \sim \mathbf{M}_7\mathbf{T}\mathbf{R}(\alpha_6)\mathbf{M}_6\mathbf{T}\mathbf{R}(\alpha_5)
\label{eq:matrix_dfn}
\end{equation}
    \item All remaining mirrors operate at near-normal incidence and the corresponding $\mathbf{M}_i$ (Eq.~\ref{eq:mirrorrefl}) are approximated as identity matrices.
 \begin{equation}
\begin{split}
\mathbf{M}_5\mathbf{T}\mathbf{R}(\alpha_4)\mathbf{M}_4\mathbf{T}\mathbf{R}(\alpha_3)\mathbf{M}_3\mathbf{T}\mathbf{R}(\alpha_2) & \sim \mathbf{T}\mathbf{R}(\alpha_4)\mathbf{T}\mathbf{R}(\alpha_3)\mathbf{T}\mathbf{R}(\alpha_2)\\
\mathbf{M}_0\mathbf{T}\mathbf{R}(\alpha_9)\mathbf{M}_9\mathbf{T}\mathbf{R}(\alpha_8)\mathbf{M}_8\mathbf{T}\mathbf{R}(\alpha_7) & \sim \mathbf{T}\mathbf{R}(\alpha_9)\mathbf{T}\mathbf{R}(\alpha_8)\mathbf{T}\mathbf{R}(\alpha_7)
\end{split}
\end{equation}
\end{enumerate}
 
Under these assumptions, the round-trip matrix reduces to
\begin{equation}
\mathbf{J}_0 \sim \mathbf{T}\mathbf{R}(\alpha_9)\mathbf{T}\mathbf{R}(\alpha_8)\mathbf{T}\mathbf{R}(\alpha_7)\mathbf{M}\mathbf{T}\mathbf{R}(\alpha_4)\mathbf{T}\mathbf{R}(\alpha_3)\mathbf{T}\mathbf{R}(\alpha_2)\mathbf{M},
\end{equation}
which can be further rearranged as:
\begin{equation}
\mathbf{J}_0 \sim \bigl( \mathbf{T}\mathbf{R}(\alpha_9)\mathbf{T} \bigr) \mathbf{R}(\alpha_8)\bigl(\mathbf{T}\mathbf{R}(\alpha_7)\mathbf{T}\bigr)\mathbf{M}\mathbf{R}(\alpha_4)\bigl(\mathbf{T}\mathbf{R}(\alpha_3)\mathbf{T}\bigr)\mathbf{R}(\alpha_2)\mathbf{M}.
\end{equation}
Noting the symmetry between the rotation angles about the bisector plane between the two wings, the commutative relationship between $\mathbf{M}$ and $\mathbf{T}$, and the identity $\mathbf{TR}(\alpha)\mathbf{T}=\mathbf{R}(-\alpha)$ \cite{Nilsson1989}, it can be shown that the round-trip matrix further reduces to:

\begin{equation}
    \mathbf{J}_0 \sim \left[\mathbf{R}(\beta)\mathbf{M}\right]^2,
\end{equation}
where $\beta = \alpha_2 - \alpha_3 + \alpha_4 = -\alpha_7 + \alpha_8 - \alpha_9$, which is the net geometric rotation accumulated per half round-trip to transform from one plane to another. We checked with our numerical simulation \cite{github} that indeed $\beta$ corresponds with a very good approximation to the angle between the two wings of the cavity, as is illustrated in Fig.~\ref{fig:points}.

The diagonalization of the matrix $[\mathbf{R}(\beta)\mathbf{M}]^2$ is detailed in Appendix B, which yields the cavity eigenvectors:
\begin{equation}
    \lvert \mathbf{u}_0^+ \rangle = \begin{pmatrix} \cos\frac{\gamma}{2}\, e^{-i\Psi/2} \\ 
    \sin\frac{\gamma}{2}\, e^{i\Psi/2} \end{pmatrix}, \quad
    \lvert \mathbf{u}_0^- \rangle = \begin{pmatrix} -\sin\frac{\gamma}{2}\, e^{-i\Psi/2} \\ 
    \cos\frac{\gamma}{2}\, e^{i\Psi/2} \end{pmatrix}
\end{equation}
with $\tan\gamma = \tan\beta/\sin(\Delta \phi)$ and $\Psi = -\left(-\Delta \phi + \frac{\pi}{2}\right)$.

The circularity of the eigenmodes, quantified by the Stokes parameter $s_3$, then takes 
the closed-form expression
\begin{equation}
    |s_3| = \left( 1+ \frac{\sin^2 (\Delta \phi)}{\tan^2 \beta}\right)^{-1/2} \left|\cos(\Delta \phi)\right|
    \label{eq:s3_analytical}
\end{equation}

\begin{figure}
    \centering
    \includegraphics[width=1.0\linewidth]{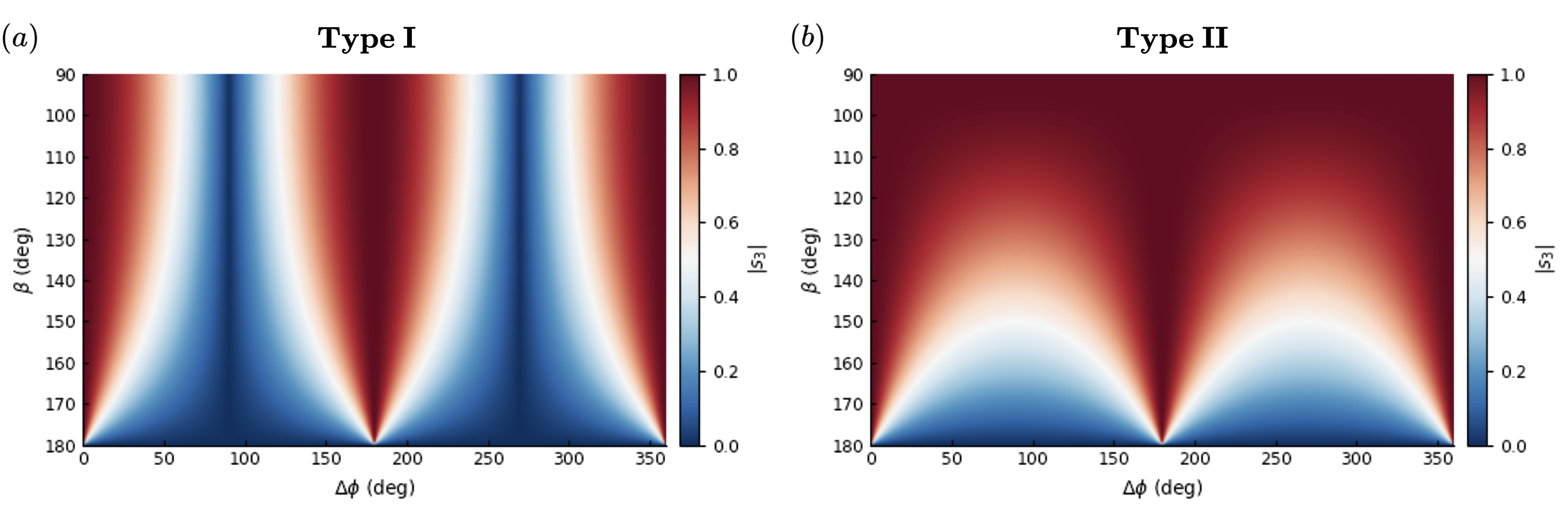}
    \caption{\textbf{Plotting of circularity as a function of the birefringence of the mirrors and the angle between the wings:} (a) Type I modes corresponding to 8 out of 10 segments of the cavity including the focusing ones. (b) Type II modes corresponding to the two outermost segments of the cavity. See Appendix \ref{sec:eigenvectors_segments}. }
    \label{fig:sim}
\end{figure}

This result is plotted in Fig.~\ref{fig:sim}, which is perfectly compatible with numerical simulations using the same assumptions and parameters. It reveals that $|s_3|$ is maximised (corresponding to circular eigenmodes) when 
$\beta = \pm\pi/2$, corresponding to the two wings of the cavity being 
arranged in orthogonal planes. To verify this, we perform a Taylor expansion around $\Delta\phi=0$ of Eq.~\eqref{eq:s3_analytical}. We obtain

\begin{equation}
    \left|s_3\right| = 1 - \frac{(\Delta\phi)^{2}}{2\sin^{2}\!\beta}
            + \mathcal{O}\!\left((\Delta\phi)^{4}\right).
    \label{eq:s3_taylor_general}
\end{equation}

For $\beta=\pi/2$, this reduces to \(\left|s_3\right| = 1 - \frac{(\Delta\phi)^2}{2}\). More generally, the factor $1/\sin^2\beta$ is minimised at $\beta=\pi/2$, so $|s_3|$ decreases more rapidly with increasing $\Delta\phi$ for any other value of $\beta$. Thus, $\beta=\pi/2$ not only maximises $|s_3|$ for a given $\Delta\phi$, but also makes it least sensitive to mirror birefringence. Although $\Delta\phi=0$ is required to reach $|s_3|=1$, operating at $\beta=\pi/2$ gives the most relaxed constraint on $\Delta\phi$ to reach a high value of circularity.

Our design satisfies this 
condition with $\alpha_2 \approx -\alpha_3 \approx \alpha_4 \approx 150^\circ$ (see \href{https://www.geogebra.org/3d/kx9vmwyp}{Geogebra}) which add up to $\beta = 90^\circ [360^\circ]$, 
simultaneously ensuring circular eigenmodes and robustness against mirror imperfections. The rest of the eigenstates in the other segments of the cavity are derived in Appendix C. In particular, we show that the situation in focusing segments $\Xi_3$ and $\Xi_8$ is the same as in $\Xi_0$.

\subsection{Numerical Simulation}

Based on the results in Sec.\ref{sec:analytical}, the orthogonal-plane Spacecraft geometry is adopted as the baseline cavity design. To assess the robustness of this geometry against realistic mirror birefringence and non-perfect symmetries, we developed a numerical code \cite{github} in parallel to this analytical model. We systematically confront the predictions from both models. The numerical simulations have also been used to make quantitative comparisons against our measurements in Sec.~\ref{sec:spacecraft_experiment}.

\section{Experimental Results} \label{sec:spacecraft_experiment}

\begin{figure}[]
    \centering
    \includegraphics[width=1.0\linewidth]{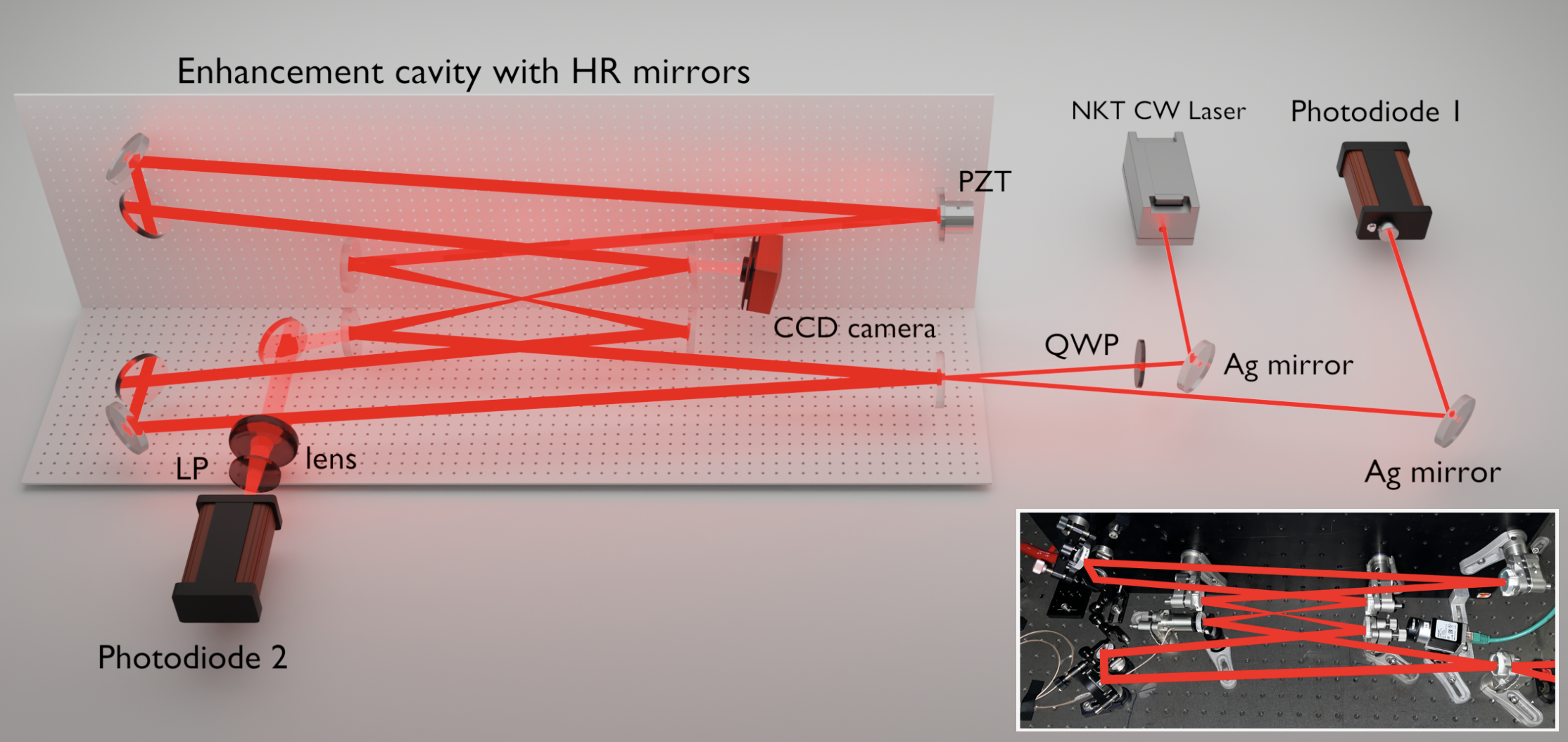}
    \caption{\textbf{Experimental Set-Up:} Injection of a CW laser at 1064 nm through a quarter wave plate (QWP) to select the circularity of input polarisation. 10-mirror cavity with a piezoelectric transducer (PZT) to sweep the cavity length. Transmission from one focusing section is viewed through a CCD camera and a photodiode, for transverse and polarisation modes measurements, respectively. Transmission from the final segment is also directed into a photodiode to observe resonance dips. \textit{Inset:} Picture of the implementation in the lab. The black breadboard is mounted vertically.}
    \label{fig:setupp}
\end{figure}

We implemented the Spacecraft cavity with orthogonal wings geometry in the laboratory, pictured in Fig.~\ref{fig:setupp}. The two focusing segments $\Xi_3$ and $\Xi_8$ are implemented with 4 Thorlabs high-reflectivity (HR) mirrors with BB1-E03 coating optimized for small angles of incidence (AOI), providing $>99.7\%$ reflectivity for both $s$ and $p$ polarizations. These are curved mirrors with focal lengths $f = +100\,\mathrm{mm}$. Each segment length is about $210\,\mathrm{mm}$ long. Four additional Edmund Optics No. 14-848 HR mirrors are used for optimizing the reflectivity for p-polarized light at $45^\circ$ AOI, they provide $>99.9\%$ reflectivity and low birefringence. Notice that, in our first attempt at implementing the Spacecraft cavity, we used Thorlabs BB1-E03 coatings for the high-AOI reflections and were unable to observe circularly polarized eigenstates. This is attributed to the large birefringence of these coatings at such high angles of incidence, illustrating that the cavity geometry alone does not eliminate the need for coatings optimized for large AOI. One flat small-AOI mirror is mounted on a piezoelectric transducer (PZT), enabling cavity length sweeping over several free spectral ranges (FSRs). Finally, the input coupler is a $99\%$ reflective mirror from Eksma Optics, which provides $1\%$ transmission into the cavity.

Real mirror birefringences were measured by polarimetry \cite{youtubepolarimetry} at their effective angle of incidence inside the cavity. Reflectivity measurements along s and p from manufacturers along with the measured birefringences are incorporated into the model. The data is summed up in Table~\ref{tab}. The observed symmetry in circularity reflects the symmetry of the mirror birefringences whereas the low deviation from zero birefringence of all mirrors results in a higher average circularity of modes.

\begin{table}[!htb]
\centering
\begin{tabular}{c c c c c c c}
\hline
Mirror $M_i$ & Mirror Type & AOI (deg) & $R_p$ & $R_s$ & $\Delta\phi$ (deg) & $s_3 = \left|\frac{S_3}{S_0}\right|$ \\
\hline
M0 & IC Eksma Optics & 5  & 0.990     & 0.990     & 0   & 0.966 \\
M1 & Edmund Optics   & 42 & 0.999   & 0.999   & 7   & 0.987 \\
M2 & Edmund Optics   & 48 & 0.999   & 0.999   & 16  & 0.980 \\
M3 & Thorlabs R=-200mm  & 8 & 0.997 & 0.997 & 12  & \textit{\textbf{0.960}} \\
M4 & Thorlabs R=-200mm  & 8 & 0.997 & 0.997 & 5   & 0.966 \\
M5 & Newport 0.5 inch & 5  & 0.999 & 0.999 & 0   & 0.966 \\
M6 & Edmund Optics   & 42 & 0.999   & 0.999   & 7   & 0.987 \\
M7 & Edmund Optics   & 48 & 0.999   & 0.999   & 16  & 0.980 \\
M8 & Thorlabs R=-200mm  & 8 & 0.997 & 0.997 & 12  & \textit{\textbf{0.960}} \\
M9 & Thorlabs R=-200mm  & 8 & 0.997 & 0.997 & 5   & 0.966 \\
\hline
\end{tabular}
\caption{Mirror parameters and degree of circular polarization, $s_3 = \left|\frac{S_3}{S_0}\right|$, in segment $\Xi_i$ between mirrors $M_i$ and $M_{i+1}$. }
\label{tab}
\end{table}

The cavity is injected with a continuous-wave NKT Koheras laser at $1064~\mathrm{nm}$ with approximately $10~\mathrm{mW}$ output power and $< 100~\mathrm{kHz}$ linewidth, its output is s-polarized. At resonance, the transmitted intracavity field from the last cavity segment interferes destructively with the reflected field from the input coupler due to their $\pi$ phase difference, producing characteristic dips in the reflected intensity. The coherent superposition of the reflected and transmitted fields is directed using a flat silver mirror onto a photodiode detector connected to an oscilloscope, where resonance features are observed during cavity length scanning. In addition, a CCD camera, and another photodiode are placed at the transmissions of the two focusing sections to observe the spatial profile and the circularity of its eigenmodes, respectively. The complete experimental setup is shown in Fig.~\ref{fig:setupp}.

\subsection{Polarization Eigenmodes Characterization}
We employ two methods to characterize the polarization eigenmodes at different cavity sections.

\subsubsection{Reflection Signal from IC}

\begin{figure}[!htb]
    \centering
    \includegraphics[width=1.0\linewidth]{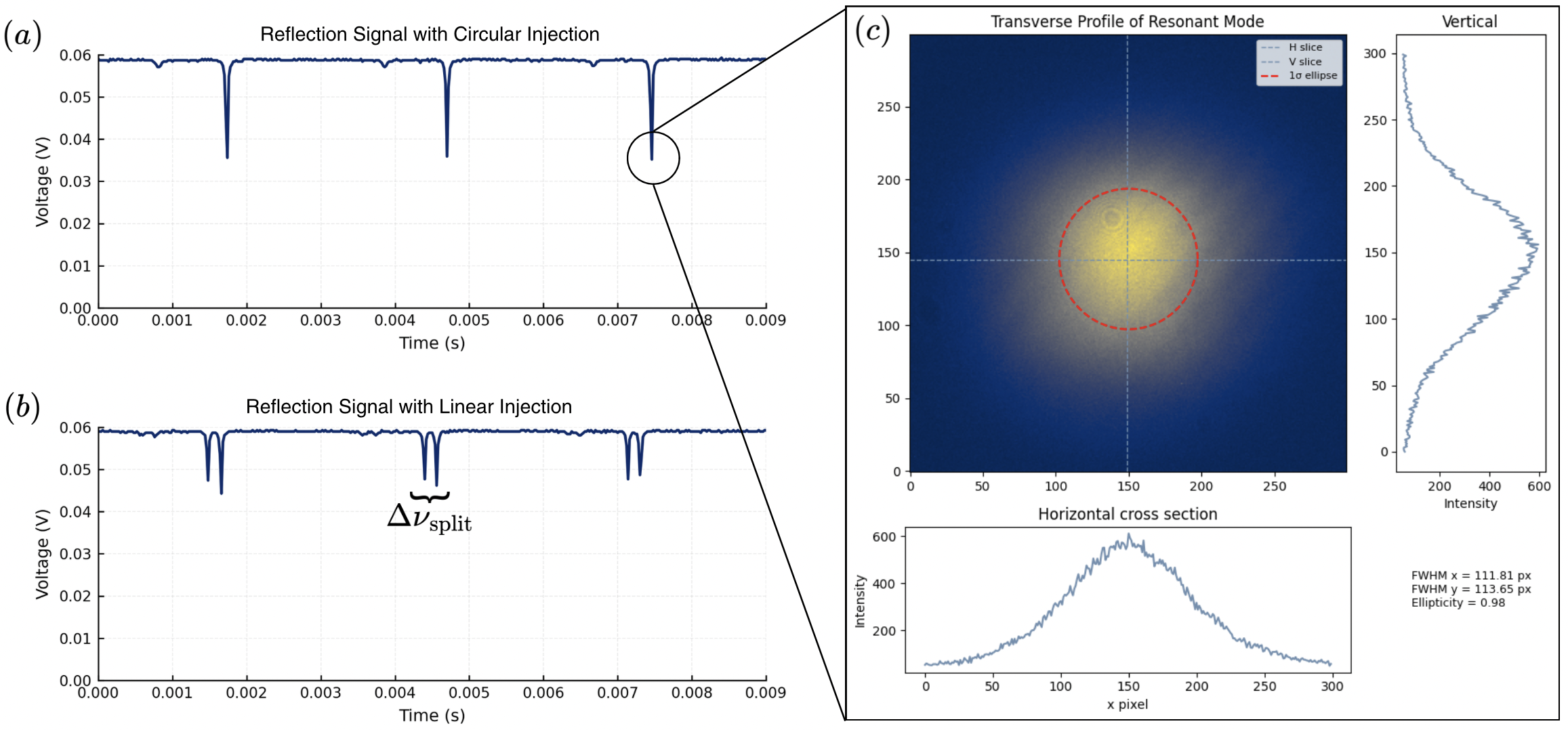}
    \caption{\textbf{Reflection signal from the input coupler as a function of the cavity length.} (a) When circularly polarized light is injected (b) When linearly polarized light is injected. (c) Beam transverse profile measured in transmission from a focusing mirror. }
    \label{fig:reflectionsig}
\end{figure}

Observing the reflected signal from the IC while sweeping the cavity length reveals a lot of information. Indeed, as stated by Eq.~(\ref{eq:resonance_matrix}), each dip in the observed signal corresponds to an eigenpolarization of the injected cavity segment. Therefore, by tuning the input polarization state one can reveal the eigenstates of the corresponding cavity segment.

The coupling efficiency to each eigenmode \(| \mathbf{u}_0^{\pm}\rangle \) depends on the overlap with the input polarization state $\mathbf{E}_{\rm inj}$:
\begin{equation}
\left| \langle \mathbf{u}_0^{\pm} | \mathbf{E}_{\rm inj} \rangle \right|^2 ,
\end{equation}

and the intensity is distributed among the eigenmodes as
\begin{equation}
\left| \langle \mathbf{u}_0^{+} | \mathbf{E}_{\rm inj} \rangle \right|^2
+
\left| \langle \mathbf{u}_0^{-} | \mathbf{E}_{\rm inj} \rangle \right|^2
= 1.
\end{equation}

If the eigenmodes are circular, linearly polarized input light couples equally to both left- and right-handed states, producing two reflection dips of equal depth. If the eigenmodes are linear, input light aligned with one eigenmode produces a single dominant dip. By rotating a quarter-wave plate (QWP) before the input coupler and observing the reflection signal, we identify the eigenmode character.

For our spacecraft cavity, we observed circular eigenmodes, as QWP orientations of ±45° allowed the input to match one circular state, completely extinguishing one resonance dip while maximizing the other, as shown in Fig.\ref{fig:reflectionsig}(a). These right and left circular polarization eigenmodes resonate at different cavity lengths due to a difference in their total phase accumulated over one round trip. The detailed derivation and the theory on the Fabry-Perot resonator is given in Appendix.\ref{sec:theory_fabry_perot}. Using Eq.(\ref{eq:resonancesplitting}), and the signal data from Fig.\ref{fig:reflectionsig}(b), the resonance splitting is 19.2°. The simulations for the same with real mirror reflectivities predict a resonance splitting of 10.2°, with the difference attributing to actual mirror birefringences deviating from those measured in Table~\ref{tab}. Nonetheless, highly circular modes in this section $\Xi_9$ suggest that the focusing segments are also likely to support circular modes, since both belong to Type I in the ideal analytical model.

\subsubsection{Polarization analysis of transmitted light from a focusing segment}

In order to further support our argument, we performed some supplementary measurements to analyse the polarization state transmitted by the end mirror of a focusing segment. The transmission signal is refocused with a lens on a photodiode to capture intensity variations as a linear polarizer is rotated (Fig.~\ref{fig:setupp}).

Since the end mirror is not at perfect normal incidence, it acts as  a linear polarizer. This effect has to be taken into account to retrieve the correct polarization state inside the focusing segment. We use the notation $(S_0, S_1, S_2, S_3)$ for the Stokes parameters inside the cavity, and $(S_0', S_1', S_2', S_3')$ for the Stokes parameters of the transmitted field. 

The measured intensity variation as a function of the angle $\theta$ of the transmission axis of the linear polarizer is given by multiplying the Stokes vector by the corresponding Mueller matrix \cite{Collett_2012}:

\begin{equation}
    I(\theta) = \tfrac{1}{2}\bigl(S_0' + S_1'\cos 2\theta + S_2'\sin 2\theta\bigr).
    \label{eq:Itheta}
\end{equation}

Note that \(S_3'\) is absent, but can be reconstructed from \(S_1'\) and \(S_2'\) since for a fully polarized state $ S_0'^2 = S_1'^2 + S_2'^2 + S_3'^2$,

\begin{equation}
    |S_3'| = \sqrt{S_0'^2 - S_1'^2 - S_2'^2} = S_0'\sqrt{1-V^2}.
    \label{eq:S3meas}
\end{equation}
where $V = \frac{I_{\max}-I_{\min}}{I_{\max}+I_{\min}}
      = \frac{\sqrt{S_1'^2+S_2'^2}}{S_0'}$

For the correction, if the p-intensity after transmission through the mirror is suppressed by a factor $\kappa$ relative to s such that $ E_s^{\rm out} = E_s^{\rm in},
    E_p^{\rm out} = \kappa \,E_p^{\rm in}$, then the corrected, normalized $S_3$ is given by:

\begin{equation}
    |s_3| = \frac{|S_3|}{S_0}
          = \frac{\kappa^{-1}\;S_0'\sqrt{1-V^2}}{\tfrac{1+1/\kappa^2}{2}\,S_0'+\tfrac{1/\kappa^2-1}{2}\,S_1'}.
\end{equation}

\begin{figure}[!htb]
    \centering
    \includegraphics[width=1.0\linewidth]{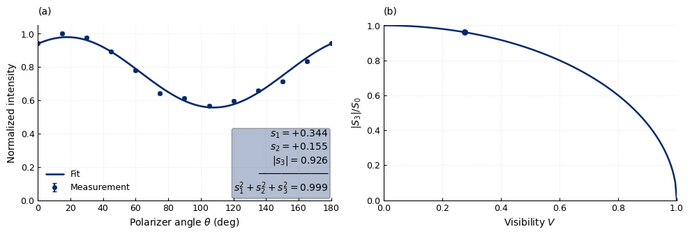}
    \caption{\textbf{Degree of Circularity from Transmisison} (a) Normalized intensity as a function of linear polarizer angle $\theta$. The corrected fit yields the normalized stokes parameters, with $|s3|=0.926$. (b) Relation between visibility and $s_3$.}
    \label{fig:trans}
\end{figure}

We measured a p suppression of 1.3 relative to s, which gives $\kappa = \frac{1}{\sqrt{1.3}}$. The final fit of the intensity fluctuations is plotted in Fig.~\ref{fig:trans}.

\subsection{Transverse Mode Characterization}

Simultaneously when the cavity is at resonance, a camera placed at the transmission captures the spatial mode profile at resonance, which exhibits a Gaussian intensity distribution, with a spatial ellipticity of 0.98, as shown in the inset in Fig.\ref{fig:reflectionsig}. The Spacecraft configuration thus supports circularly polarized eigenmodes, while its symmetry also eliminates most of the spatial astigmatism \cite{Winkler:16} despite accommodating large angles of incidence.

\section{Extension of the design for CE-HHG implementation} \label{sec:femtosecond_enhancement}

Extending the Spacecraft cavity to a femtosecond enhancement cavity requires broadband mirror coatings satisfying $\Delta\phi_{i,\omega} \to 0$ across the full driving bandwidth, that is, near-zero GDD and birefringence along with a high LIDT to withstand the circulating fluence. A target enhancement of $\sim 500$ requires a finesse of $\sim 1000$. In the all-reflective design, the high-AOI mirrors are the dominant source of losses as they are typically less efficient than those at small AOI. Hence, these mirrors must be prioritized in coating optimization.

The current implementation of the Spacecraft cavity has a round-trip length of $\sim 4\,\mathrm{m}$, which can be extended to $6\,\mathrm{m}$ or beyond either by translating the six outer folding mirrors within their plane or by inserting additional symmetric mirror pairs at small AOI, in both cases without altering the fundamental symmetry or polarization eigenmode structure. Achieving this in practice places stringent simultaneous demands on reflectivity, GDD, birefringence, and LIDT from mirror manufacturers, while requiring the geometry to be implemented precisely as described here.

\section{Conclusion} \label{sec:conclusion}
In this work, we proposed and demonstrated the \textit{Spacecraft} cavity, a novel all-reflective optical cavity designed to implement non-collinear circularly polarized high harmonic generation. The proposed geometry satisfies the constraints required for NCP-HHG, while analytical and numerical models show that orthogonal cavity planes maximise the circularity and robustness of the eigenmodes against mirror birefringence. Experimental characterization of a continuous-wave prototype confirms the existence of fairly circular polarization eigenmodes in agreement with the theoretical predictions. Thus, this cavity provides a practical route towards implementing NCP-HHG inside femtosecond cavities, and future work will focus on implementing this and potentially achieving more efficient XUV output coupling of all the harmonics.

\section{Acknowledgements} \label{sec:conclusion}

This research received the financial support of the
French National Research Agency through Grants No.
ANR-24-CE30-2261-ASAP. We acknowledge funding from the
CEA-Audace! Program under ANR-France 2030 No.
ANR24-RRII-0004. We thank C. Blondel and C. Drag for lending us some equipment, and C. Heyl for fruitful discussions.


\bibliography{references}

\begin{thebibliography}{10}
\newcommand{\enquote}[1]{``#1''}

\bibitem{Hickstein_Dollar_Grychtol_Ellis_Knut_Hernández-García_Zusin_Gentry_Shaw_Fan_et_al._2015}
D.~D. Hickstein, F.~J. Dollar, P.~Grychtol, \emph{et~al.},
  \enquote{Non-collinear generation of angularly isolated circularly polarized
  high harmonics,} {\protect\JournalTitle{Nature Photonics}} \textbf{9},
  743–750 (2015).

\bibitem{Ferray_L’Huillier_Li_Lompre_Mainfray_Manus_1988}
M.~Ferray, A.~L’Huillier, X.~F. Li, \emph{et~al.}, \enquote{Multiple-harmonic
  conversion of 1064 nm radiation in rare gases,}
  {\protect\JournalTitle{Journal of Physics B: Atomic, Molecular and Optical
  Physics}} \textbf{21}, L31 (1988).

\bibitem{Gohle_Udem_Herrmann_Rauschenberger_Holzwarth_Schuessler_Krausz_Hänsch_2005}
C.~Gohle, T.~Udem, M.~Herrmann, \emph{et~al.}, \enquote{A frequency comb in the
  extreme ultraviolet,} {\protect\JournalTitle{Nature}} \textbf{436}, 234–237
  (2005).

\bibitem{Jones_Moll_Thorpe_Ye_2005}
R.~J. Jones, K.~D. Moll, M.~J. Thorpe, and J.~Ye, \enquote{Phase-coherent
  frequency combs in the vacuum ultraviolet via high-harmonic generation inside
  a femtosecond enhancement cavity,} {\protect\JournalTitle{Physical Review
  Letters}} \textbf{94}, 193201 (2005).

\bibitem{Allison2025}
T.~K. Allison, A.~Kunin, and G.~Schönhense, \enquote{Cavity-enhanced
  high-order harmonic generation for high-performance time-resolved
  photoemission experiments,} {\protect\JournalTitle{APL Photonics}}
  \textbf{10}, 010906 (2025).

\bibitem{Zhang2024}
C.~Zhang, T.~Ooi, J.~S. Higgins, \emph{et~al.}, \enquote{Frequency ratio of the
  229mth nuclear isomeric transition and the 87sr atomic clock,}
  {\protect\JournalTitle{Nature}} \textbf{633}, 63--70 (2024).

\bibitem{Pupeza_Zhang_Högner_Ye_2021}
I.~Pupeza, C.~Zhang, M.~Högner, and J.~Ye, \enquote{Extreme-ultraviolet
  frequency combs for precision metrology and attosecond science,}
  {\protect\JournalTitle{Nature Photonics}} \textbf{15}, 175–186 (2021).

\bibitem{Zhang_Schoun_Heyl_Porat_Gaarde_Ye_2020}
C.~Zhang, S.~B. Schoun, C.~M. Heyl, \emph{et~al.}, \enquote{Noncollinear
  enhancement cavity for record-high out-coupling efficiency of an extreme-uv
  frequency comb,} {\protect\JournalTitle{Physical Review Letters}}
  \textbf{125}, 093902 (2020).

\bibitem{PhysRevResearch.7.023071}
S.~H. Wissenberg, J.~Weitenberg, J.~Moreno, \emph{et~al.},
  \enquote{Noncollinear enhancement resonator with intrinsic pulse
  synchronization and alignment employing wedge mirrors,}
  {\protect\JournalTitle{Phys. Rev. Res.}} \textbf{7}, 023071 (2025).

\bibitem{Chappuis2019}
C.~Chappuis, D.~Bresteau, T.~Auguste, \emph{et~al.}, \enquote{High-order
  harmonic generation in an active grating,} {\protect\JournalTitle{Phys. Rev.
  A}} \textbf{99}, 033806 (2019).

\bibitem{Jones_1941}
R.~C. Jones, \enquote{A new calculus for the treatment of optical systemsi
  description and discussion of the calculus,} {\protect\JournalTitle{Journal
  of the Optical Society of America}} \textbf{31}, 488 (1941).

\bibitem{Chow_Gea-Banacloche_Pedrotti_Sanders_Schleich_Scully_1985}
W.~W. Chow, J.~Gea-Banacloche, L.~M. Pedrotti, \emph{et~al.}, \enquote{The ring
  laser gyro,} {\protect\JournalTitle{Reviews of Modern Physics}} \textbf{57},
  61–104 (1985).

\bibitem{Nilsson1989}
A.~Nilsson, E.~Gustafson, and R.~Byer, \enquote{Eigenpolarization theory of
  monolithic nonplanar ring oscillators,} {\protect\JournalTitle{IEEE Journal
  of Quantum Electronics}} \textbf{25}, 767--790 (1989).

\bibitem{Zomer_Fedala_Pavloff_Soskov_Variola_2009}
F.~Zomer, Y.~Fedala, N.~Pavloff, \emph{et~al.}, \enquote{Polarization induced
  instabilities in external four-mirror fabry-perot cavities,}
  {\protect\JournalTitle{Applied Optics}} \textbf{48}, 6651–6661 (2009).

\bibitem{geogebraGeoGebraWorlds}
\enquote{{G}eo{G}ebra - the world’s favorite, free math tools used by over
  100 million students and teachers --- geogebra.org,}
  \url{https://www.geogebra.org/}.

\bibitem{reZonator}
\url{http://www.rezonator.orion-project.org}.

\bibitem{github}
\url{https://github.com/Attolab/cavities}.

\bibitem{youtubepolarimetry}
\enquote{Thorlabs video tutorial "build a polarimeter",}
  \url{https://www.youtube.com/watch?v=pR4r7gMyN5U&list=PLN3i-H51ZELgsYOSg0f_n8gqYgz2AlFha&index=5}.

\bibitem{Collett_2012}
E.~Collett, \emph{Field Guide to Polarization}, SPIE Field Guides (SPIE Press,
  Bellingham, WA, 2012), 3rd ed. See pp.~23--25.

\bibitem{Winkler:16}
G.~Winkler, J.~Fellinger, J.~Seres, \emph{et~al.}, \enquote{Non-planar
  femtosecond enhancement cavity for vuv frequency comb applications,}
  {\protect\JournalTitle{Opt. Express}} \textbf{24}, 5253--5262 (2016).

\bibitem{cohen1977mq1}
C.~Cohen-Tannoudji, B.~Diu, and F.~Laloë, \emph{Mécanique quantique. Tome I},
  Collection Enseignement des sciences (Hermann, Paris, 1977), nouvelle
  édition revue, corrigée et augmentée ed. See Complément B IV, p.~418.

\bibitem{polstokes}
H.~Berry, G.~Gabrielse, and A.~Livingston, \enquote{Measurement of the stokes
  parameters of light,} {\protect\JournalTitle{Applied Optics}} \textbf{16},
  3200--3205 (1977).

\end{thebibliography}






\appendix

\section{Theory of the Fabry-Perot resonator} \label{sec:theory_fabry_perot}

In this appendix, we review the Fabry-Perot resonator theory and establish that each eigenvector of a roundtrip matrix for a particular segment is the polarization state associated to a particular resonant condition. We tightly follow the demonstration by Zomer and coauthors \cite{Zomer_Fedala_Pavloff_Soskov_Variola_2009}.

Let $\{ \lvert \mathbf{u}_{0}^{\pm} \rangle \}$ be the basis of eigenvectors of $\mathbf{J}_0$. The +/- notation refering to the north and south hemispheres of the Poincaré sphere. In this basis, the matrix is diagonal and can be expressed as

\begin{equation}
\mathbf{U}_0^{-1} \mathbf{J}_0 \mathbf{U}_0
=
\begin{pmatrix}
\mu^{+} & 0 \\
0 & \mu^{-} \\
\end{pmatrix},
\end{equation}

with $\mu^{\pm}$ the eigenvalues of the $\mathbf{J}_0$ matrix, and

\begin{equation}
\mathbf{U}_0
=
\begin{pmatrix}
\mathbf{u}_0^{+}.\mathbf{p}_0^{\text{out}} & \mathbf{u}_0^{-}.\mathbf{p}_0^{\text{out}} \\
\mathbf{u}_0^{+}.\mathbf{s}_0^{\text{out}} & \mathbf{u}_0^{-}.\mathbf{s}_0^{\text{out}} \\
\end{pmatrix}.
\end{equation}

Notice that we do not need to subscript the eigenvalues, since they do not depend on the segment. Indeed, all the roundtrip matrices have the same eigenvalues, as demonstrated with Eq. (\ref{eq:transformeigvecs}). However they can have different eigenvectors.

\begin{figure}[!htb]
    \centering
    \includegraphics[width=0.6\linewidth]{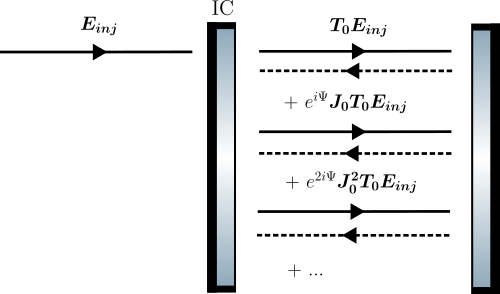}
    \caption{\textbf{Coherent superposition}}
    \label{fig:coherentsuperposition}
\end{figure}

The electric field in the $\Xi_0$ segment can be calculated \cite{Zomer_Fedala_Pavloff_Soskov_Variola_2009} as the coherent sum of the fields resulting from successive roundtrips, as is illustrated in Fig.~\ref{fig:coherentsuperposition}. Explicitely in the $B_0^{\text{out}}$ basis:

\begin{equation}
\begin{pmatrix}
E_{0}^{\text{out},p} \\
E_{0}^{\text{out},s}
\end{pmatrix}
= \left[ \sum_{n=0}^{\infty} \left(e^{i\Psi} \mathbf{J}_0  \right)^n \right] \mathbf{T}_{0}
\begin{pmatrix}
E_{\rm inj}^{p} \\
E_{\rm inj}^{s}
\end{pmatrix}
\label{eq:circulatingfield}
\end{equation}

with $\mathbf{E}_{\rm inj}$ the injected field expressed in the $B_{0}^{\text{out}}$ coordinate system, $\mathbf{T}_0$ the transmission matrix of the IC, $\Psi = 2\pi \Lambda/\lambda$ is the roundtrip phase independant of $\phi_i^p$ and $\phi_i^s$ that we can think of as accounting for the free space propagation over the cumulated distance between mirrors $\Lambda$. This distance is precisely controllable experimentally by mounting a mirror on a piezoelectric actuator (PZT).

For simplicity, we will consider that the transmission matrix $\mathbf{T}_0$ of the IC is the identity, which is a good approximation if the IC is close to normal incidence. In transmission, no birefringence can be observed as long as the substrate and the coating are isotropic, which is our case. However, the polarization state will be affected in transmission as a result of losses anisotropy, which could be strong with a high AOI. This could be precompensated, but would add some complexity in the optical system at injection.

We now rewrite Eq. \ref{eq:circulatingfield} in the $\{ \mathbf{u}_0^{+}, \mathbf{u}_0^{-} \}$ basis:

\begin{equation}
\begin{pmatrix}
E_{0}^{+} \\
E_{0}^{-}
\end{pmatrix}
= \left[ \sum_{n=0}^{\infty} \left(e^{i\Psi} \mathbf{U}_0^{-1} \mathbf{J}_0 \mathbf{U}_0 \right)^n \right]
\begin{pmatrix}
E_{\rm inj}^{+} \\
E_{\rm inj}^{-}
\end{pmatrix},
\label{eq:circulatingfield+}
\end{equation}

with

\begin{equation}
\begin{pmatrix}
E_0^{+} \\
E_0^{-}
\end{pmatrix}
= \mathbf{U}_0
\begin{pmatrix}
E_0^{\text{out}, p} \\
E_0^{\text{out}, s}
\end{pmatrix}
\,\text{and}\,
\begin{pmatrix}
E_{\rm inj}^{+} \\
E_{\rm inj}^{-}
\end{pmatrix}
= \mathbf{U}_0
\begin{pmatrix}
E_{\rm inj}^{p} \\
E_{\rm inj}^{s}
\end{pmatrix}.
\end{equation}

Using

\begin{equation}
\sum_{n=0}^{\infty}
\begin{pmatrix}
\mu^{+ n} e^{i n \Psi} & 0 \\
0 & \mu^{- n} e^{i n \Psi} \\
\end{pmatrix}
=
\begin{pmatrix}
\frac{1}{1- \mu^{+} e^{i\Psi}} & 0 \\
0 & \frac{1}{1- \mu^{-} e^{i\Psi}} \\
\end{pmatrix},
\end{equation}

we end up with

\begin{equation}
\begin{pmatrix}
E_0^{+} \\
E_0^{-}
\end{pmatrix}
=
\begin{pmatrix}
\frac{1}{1- \lvert \mu^+ \rvert e^{i (\arg{\mu^+} + \Psi)}} & 0 \\
0 & \frac{1}{1- \lvert \mu^- \rvert e^{i (\arg{\mu^-} + \Psi)}} \\
\end{pmatrix}
\begin{pmatrix}
E_{\rm inj}^{+} \\
E_{\rm inj}^{-}
\end{pmatrix}.
\label{eq:resonance_matrix}
\end{equation}

This last expression evokes two resonances - one for each eigenpolarization of the roundtrip matrix - for two different total cavity lengths $\Lambda^{\pm}$ that are solutions of the equations

\begin{equation}
\arg{\mu^{\pm}} + 2\pi \frac{\Lambda^{\pm}}{\lambda} = 0 [2 \pi],
\end{equation}

which leads to

\begin{equation}
\frac{\Lambda^{+} - \Lambda^{-}}{\lambda} = - \frac{\arg{\mu^{+}} - \arg{\mu^{-}}}{2\pi}.
\end{equation}

A variation of the total length of the cavity is related to a variation of the resonant frequency by

\begin{equation}
\frac{\Delta \nu}{\Delta \nu_{\text{FSR}}} = - \frac{\Delta \Lambda}{\lambda}.
\end{equation}

Therefore, we end up with a frequency splitting that is proportional to the difference of the arguments of the eigenvalues of the round trip matrix

\begin{equation}
\frac{\Delta \nu_{\text{split}}}{\Delta \nu_{\text{FSR}}} = \frac{\arg{\mu^{+}} - \arg{\mu^{-}}}{2\pi}
\label{eq:resonancesplitting}
\end{equation}

Note that the resonance condition can equally be satisfied by varying the injected laser frequency while keeping the total cavity length fixed. This is the natural picture for the frequency comb injection, where the cavity offset frequency is given by $\arg\mu^{\pm}/(2\pi) \cdot \Delta\nu_{\text{FSR}}$.

\section{Diagonalization of the Reduced Round-Trip Matrix}
We are interested in the eigenvectors and the stokes parameter $s_3$ of the matrix $[\mathbf{R}(\beta)\mathbf{M}]^2$. Note that since it has the same eigenvectors as $\mathbf{R}(\beta)\mathbf{M}$, it is sufficient to diagonalize $\mathbf{N}=\mathbf{R}(\beta)\mathbf{M}$ with
\[
\mathbf{M} =
\begin{pmatrix}
e^{i\Delta \phi} & 0 \\
0 & e^{-i\Delta \phi}
\end{pmatrix},
\]

defined in Eq. (\ref{eq:matrix_dfn}) with $\Delta \phi$ the birefringence of a single mirror. Note that the expression of $\mathbf{M}$ takes into account two successive reflections, so the total birefringence is $2\Delta \phi$.

We write explicitly
\begin{equation}
\mathbf{N} =
\begin{pmatrix}
e^{i\Delta \phi}\cos\beta & -e^{-i\Delta \phi}\sin\beta \\
e^{i\Delta \phi}\sin\beta & e^{-i\Delta \phi}\cos\beta
\end{pmatrix}.
\end{equation}

We compute
\begin{equation}
\begin{split}
\frac{1}{2}(N_{11}+N_{22}) &= \cos\beta \cos(\Delta \phi), \\
\frac{1}{2}(N_{11}-N_{22}) &= i\,\cos\beta \sin(\Delta \phi).
\end{split}
\end{equation}

We decompose $\mathbf{N}$ as $\mathbf{I}$ and $\mathbf{K}$, where $\mathbf{K}$ is Hermitian and can be diagonalised \cite{cohen1977mq1}

\begin{equation}
\mathbf{N} = \cos\beta \cos(\Delta \phi)\,\mathbf{I}
+ i\,\cos\beta \sin(\Delta \phi)\,\mathbf{K},
\end{equation}
with
\begin{equation}
\mathbf{K} =
\begin{pmatrix}
1 & \dfrac{i e^{-i\Delta \phi}\tan\beta}{\sin(\Delta \phi)} \\
\dfrac{-i e^{i\Delta \phi}\tan\beta}{\sin(\Delta \phi)} & -1
\end{pmatrix}.
\end{equation}

We define $\gamma \in [0,\pi]$ such that
\begin{equation}
\tan\gamma = \frac{2|K_{21}|}{K_{11}-K_{22}} = \frac{\tan\beta}{\sin(\Delta \phi)},
\end{equation}
and
\begin{equation}
\Psi = -\left(-\Delta \phi + \frac{\pi}{2}\right).
\end{equation}

We rewrite $\mathbf{K}$ as
\begin{equation}
\mathbf{K} =
\begin{pmatrix}
1 & e^{-i\Psi}\tan\gamma \\
e^{i\Psi}\tan\gamma & -1
\end{pmatrix}.
\end{equation}

The eigenvectors of $\mathbf{K}$, and therefore of $\mathbf{N}$ and $[\mathbf{R}(\beta)\mathbf{M}]^2$, are \cite{cohen1977mq1}
\begin{equation}
    \lvert \mathbf{u}_0^+ \rangle = \begin{pmatrix} \cos\frac{\gamma}{2}\, e^{-i\Psi/2} \\ 
    \sin\frac{\gamma}{2}\, e^{i\Psi/2} \end{pmatrix} \quad 
    \lvert \mathbf{u}_0^- \rangle = \begin{pmatrix} -\sin\frac{\gamma}{2}\, e^{-i\Psi/2} \\ 
    \cos\frac{\gamma}{2}\, e^{i\Psi/2} \end{pmatrix}
\label{eigvecs0}
\end{equation}

Using Eq.~(\ref{eq:s3_expression}), we obtain

\begin{equation}
|s_3| = |\sin\gamma \sin\Psi| = |\sin\gamma \cos(\Delta \phi)|.
\end{equation}

Using
\[\sin\gamma = \sin\left[\arctan\left(\frac{\tan\beta}{\sin(\Delta \phi)}\right)\right],\]
and $\left|\sin(\arctan (x))\right| = (1 + 1/x^2)^{-1/2}$, we get

\begin{equation}
|s_3| = \left( 1+ \frac{\sin^2 (\Delta \phi)}{\tan^2 \beta}\right)^{-1/2} \left|\cos(\Delta \phi)\right|.
\end{equation}

\section{Derivation of the Polarization States in All Segments}
\label{sec:eigenvectors_segments}

From the expression of the round-trip matrix in Eq.~(\ref{eq:roundtrip_matrix}), one can deduce the following recurrence relation between the eigenstates of successive segments:

\begin{equation}
    \mathbf{J}_{i+1}[\mathbf{M}_{i+1}\mathbf{T}\mathbf{R}_{i}\lvert \mathbf{u}_i^\pm \rangle] = \mathbf{M}_{i+1}\mathbf{T}\mathbf{R}_{i}\mathbf{J}_{i}\lvert \mathbf{u}_i^\pm \rangle = \mu^\pm[\mathbf{M}_{i+1}\mathbf{T}\mathbf{R}_i\lvert \mathbf{u}_i^\pm \rangle]
    \label{eq:transformeigvecs}
\end{equation}

Therefore one can directly deduce the eigenvectors of segment $\Xi_{i+1}$ from those in segment $\Xi_i$ with the formula

\begin{equation}
\lvert \mathbf{u}_{i+1}^\pm \rangle = \mathbf{M}_{i+1}\mathbf{T}\mathbf{R}_{i}\lvert \mathbf{u}_i^\pm \rangle
\label{eq:recurrence_vecs}
\end{equation}

which simply means that the relation between the polarization states in successive segments in the resonating case is the same as it would be in free space.

From the eigenvectors derived for segment $\Xi_0$ in Eq.~(\ref{eq:recurrence_vecs}), we use the identity in Eq.~(\ref{eq:transformeigvecs}) to obtain the eigenvectors in other segments to get the expression of $|s_3|$ as a function of non-planarity angle $\beta$ and mirror birefringence $\Delta\phi$.

In the segment $\Xi_1$, since $\mathbf{R}_0 = \mathbf{I}$, the eigenstates are:

\begin{equation}
\lvert \mathbf{u}_1^+ \rangle = \mathbf{M}_1\mathbf{T}\lvert \mathbf{u}_0^+ \rangle = \begin{pmatrix} -\cos\frac{\gamma}{2}\, e^{i(-\Psi/2 + \Delta \phi/2)} \\ 
\sin\frac{\gamma}{2}\, e^{i(\Psi/2 - \Delta \phi/2)} \end{pmatrix} \quad
\lvert \mathbf{u}_1^- \rangle = \mathbf{M}_1\mathbf{T}\lvert \mathbf{u}_0^- \rangle = \begin{pmatrix} \sin\frac{\gamma}{2}\, e^{i(-\Psi/2 + \Delta \phi/2)} \\ 
\cos\frac{\gamma}{2}\, e^{i(\Psi/2 - \Delta \phi/2)} \end{pmatrix}
\end{equation}

And following similar steps as before, we obtain for $\Xi_1$:
\begin{equation}
\boxed{
|s_3| = \left( 1+ \frac{\sin^2 (\Delta \phi)}{\tan^2 \beta}\right)^{-1/2}
}
\end{equation}

Similar transformations can be performed for other segments and the resulting expressions for $|s_3|$ are summarised in Table~\ref{tabpolarization}.

\begin{table}[!htb]
\centering
\begin{tabular}{c c c}
\hline
\textbf{Circularity Type} & \textbf{Segments} & \textbf{Expression for $|s_3|$} \\
\hline
Type I & $\Xi_0$, $\Xi_2$, $\Xi_3$, $\Xi_4$, $\Xi_5$, $\Xi_7$, $\Xi_8$, $\Xi_9$ & $ \left( 1+ \frac{\sin^2 (\Delta \phi)}{\tan^2 \beta}\right)^{-1/2}|\cos(\Delta\phi)|$ \\
Type II & $\Xi_1$, $\Xi_6$ & $ \left( 1+ \frac{\sin^2 (\Delta \phi)}{\tan^2 \beta}\right)^{-1/2}$ \\
\end{tabular}
\caption{Circularity type for each segment.}
\label{tabpolarization}
\end{table}

The focusing segments that are of primary interest in this analysis are {$\Xi_3$} and {$\Xi_8$} and both of these segments follow Type I circularity relationship.

\section{Polarization analysis from the transmitted light of a focusing segment}
\label{sec:appendix_stokes}

A dielectric mirror under non-normal transmission acts as a partial linear polarizer, so this effect must be accounted for in order to reconstruct the true intracavity polarization state. 

A linear polarizer with transmission axis at angle \( \theta \) to the s axis of the mirror gives an intensity function \cite{polstokes} as follows:

\begin{equation}
    I(\theta) = \tfrac{1}{2}\bigl(S_0' + S_1'\cos 2\theta + S_2'\sin 2\theta\bigr).
    \label{eq:Itheta}
\end{equation}

Note that \(S_3'\) is absent, but can be reconstructed from \(S_1\) and \(S_2\).

The above equation can be written with 3 parameters:
\begin{equation}
    I(\theta) = A + B\cos 2\theta + C\sin 2\theta,
\end{equation}
which is fitted directly to the measured data \(\{(\theta_i, I_i)\}\). The fit parameters map onto the measured Stokes parameters as:
\begin{equation}
    S_0' = 2A, \qquad S_1' = 2B, \qquad S_2' = 2C.
    \label{eq:fit2stokes}
\end{equation}

Note also that these fit parameters are also related to the max and min of the intensity curve as:
\begin{equation}
    A = \tfrac{I_{\max}+I_{\min}}{2}, \qquad
    \sqrt{B^2+C^2} = \tfrac{I_{\max}-I_{\min}}{2},
\end{equation}
so the \textit{visibility} of the fringe pattern is
\begin{equation}
    V = \frac{I_{\max}-I_{\min}}{I_{\max}+I_{\min}}
      = \frac{\sqrt{B^2+C^2}}{A}
      = \frac{\sqrt{S_1'^2+S_2'^2}}{S_0'}.
    \label{eq:visibility}
\end{equation}

Since the state is fully polarized, it lies on the Poincaré sphere:
\begin{equation}
    S_0'^2 = S_1'^2 + S_2'^2 + S_3'^2.
\end{equation}
Combining with Eq.~\eqref{eq:visibility}:
\begin{align}
    |S_3'| &= \sqrt{S_0'^2 - S_1'^2 - S_2'^2} \notag\\
            &= S_0'\sqrt{1-V^2}.
    \label{eq:S3meas}
\end{align}

The sign of $S_3'$ cannot be determined but it is redundant since we know that the cavity holds two orthogonal eigenmodes with same $|s_3'|$ but opposite signs.

We now take into account that the transmission through the end mirror with an incidence angle about 8° acts as a partial polarizer. In this case, we measured that the transmitted intensity of p-polarized light is 1.3 times the one from s-polarized light, which can be translated in amplitude as:
\begin{equation}
    E_s' = \kappa E_s, \qquad E_p' = E_p 
\end{equation}

with $\kappa = 1 /\sqrt{1.3}$.
Writing the Stokes parameters in terms of field amplitudes:

\begin{align}
    S_0 &= |E_s|^2 + |E_p|^2, &
    S_0' &= \kappa^2|E_s|^2 + |E_p|^2,\\
    S_1 &= |E_s|^2 - |E_p|^2, &
    S_1' &= \kappa^2|E_s|^2 - |E_p|^2,\\
    S_2 &= 2\,\mathrm{Re}(E_s^*E_p), &
    S_2' &= 2\,\mathrm{Re}(\kappa E_s^*E_p) = \kappa S_2,\\
    S_3 &= 2\,\mathrm{Im}(E_s^*E_p), &
    S_3' &= 2\,\mathrm{Im}(\kappa E_s^*E_p) = \kappa S_3.
\end{align}

Solving for the true Stokes parameters from the measured ones:

\begin{equation}
    \begin{aligned}
        S_0 &= 1.15\,S_0' + 0.15\,S_1'\\
        S_1 &= 0.15\,S_0' + 1.15\,S_1'\\
        S_2 &= \sqrt{1.3}\;S_2'\\
        |S_3| &= \sqrt{1.3}\;|S_3'|
    \end{aligned}
    \label{eq:correction}
\end{equation}

where the numerical coefficients come from
\(\tfrac{1+1/\kappa^2}{2} = \tfrac{1+1.3}{2} = 1.15\) and
\(\tfrac{1/\kappa^2-1}{2} = 0.15\).

The important parameters are the normalised Stokes \(s_i = S_i/S_0\), which live on the unit Poincaré sphere, for which 
 both the numerator and denominator must be corrected:

\begin{equation}
    s_1 = \frac{S_1}{S_0} = \frac{0.15\,S_0'+1.15\,S_1'}{1.15\,S_0'+0.15\,S_1'},
\end{equation}
\begin{equation}
    s_2 = \frac{S_2}{S_0} = \frac{\sqrt{1.3}\;S_2'}{1.15\,S_0'+0.15\,S_1'},
\end{equation}
\begin{equation}
    |s_3| = \frac{|S_3|}{S_0}
          = \frac{\sqrt{1.3}\;S_0'\sqrt{1-V^2}}{1.15\,S_0'+0.15\,S_1'}.
    \label{eq:s3norm}
\end{equation}

Finally, to verify the fit of our parameters, for a fully polarized state the purity check must satisfy:
\begin{equation}
    s_1^2 + s_2^2 + s_3^2 = 1.
\end{equation}

\end{document}